# A Rich-Variant Architecture for a User-Aware multi-tenant SaaS approach

Houda Kriouile[1] and Bouchra El Asri[2]

[1] IMS Team, ADMIR Lab, ENSIAS, Rabat IT Center, Mohammed V University in Rabat
Rabat, Morocco

[2] IMS Team, ADMIR Lab, ENSIAS, Rabat IT Center, Mohammed V University in Rabat
Rabat, Morocco

**Abstract**

Software as a Service cloud computing model favorites the Multi-Tenancy as a key factor to exploit economies of scale. However Multi-Tenancy present several disadvantages. Therein, our approach comes to assign instances to multi-tenants with an optimal solution while ensuring more economies of scale and avoiding tenants hesitation to share resources. The present paper present the architecture of our user-aware multi-tenancy SaaS approach based on the use of rich-variant components. The proposed approach seek to model services functional customization as well as automation of computing the optimal distribution of instances by tenants. The proposed model takes into consideration tenants functional requirements and tenants deployment requirements to deduce an optimal distribution using essentially a specific variability engine and a graph-based execution framework.

*Keywords: Cloud Computing, SaaS, Multi-Tenancy, Rich-Variant Component, Rich-Variant Architecture.*

## 1. Introduction

The cloud computing idea dates back to 1961 [1] and it has become a true technological trend, able to carry out the strategies of companies in terms of optimization and rationalization of expenses related to IT. Nowadays, cloud computing is one of the most used technologies for building and delivering IT services, using different service delivery models depending on services nature.

Cloud computing characteristics - as on-demand self-service, wide network access, resource pooling, fast elasticity, and measurable service - enable to give the illusion of having infinite resources available, to request when and as we want. Thus, it is more interesting to have a software as a service from a cloud provider, than to build an entire datacenter, with the overcrowding of the components needed to make the same service available internally.

However, cloud computing adoption is not so intuitive and so encouraging that it seems. Indeed, technical, organizational and economic obstacles make the decision to adopt cloud computing critical and hesitant. The scientific community [2][3] has a great interest in this area: Indeed, several research works focuses on the solution proposal for each obstacle encountered.

It is in this context that our RV-Cloud approach takes place where we are interested in one of the commonly accepted Cloud service delivery models, namely Software as a Service (SaaS) which refers to a software distribution model wherein the applications are hosted by a service provider and made available to clients on a network. In particular, we seek to provide SaaS providers with a more flexible, reusable and dynamic system, while allowing them more economy and less service customers reluctance.

As a key factor in exploiting economies of scale, SaaS favors the Multi-Tenancy (MT), a notion of sharing resources within a large group of customer organizations, called tenants. While MT brings several benefits to SaaS, however, it only meets the requirements that are common to all tenants. In addition, tenants themselves are hesitant about tenancy sharing especially that they need applications variability management to meet their specific needs. And on the other hand, they have fears of disclosure of their information with other tenants, competitors for example.

Thus, in order to provide elements of answer to the problem of the variability management of SaaS applications, several research works were carried out to propose approaches focusing on the facilitation of customization of SaaS applications according to tenants specific requirements [4] [5] [6] [7]. These works are generally based on the exploitation of MT advantages, applications variability management mechanisms, and tenants isolation on the same instance.

Similarly, our approach aims to create a flexible and reusable environment that allows greater flexibility and elasticity for customers while taking advantage from economies of scale. The proposed approach is a user-aware solution that integrates a functional variability across application components and a deployment variability at multi-tenant level. In addition, the approach focuses on satisfying both stakeholders, providers and customers, while maintaining a level of performance and efficiency.

To meet the need for reuse and flexibility, we take advantage of the paradigm of component-based system development that brings several benefits such as an improved reuse, a huge flexibility, a configurability, and a better scalability. These systems are usually built from components whose individual behavior is well known and correct. Moreover, through a combination of multi-functionality and MT, we seek to benefit from the multifunctional concept of multiviews as well as the high configurability feature of MT, in order to allow some economies of scale for SaaS application providers while minimizing the cost for customers tenants of its applications. We aim to achieve our goals by using Rich-Variant Component (RVC) that provide more sharing capabilities allowing more instance-sharing, more cost reduction, as well as better communication between tenants communities. Besides, we use the basics and some theorems of graph theory to find the optimal distribution of RVC instances on tenants.

The objective of our approach is to assign instances to different tenants with a solution using a less number of instances, thus a more optimal solution than existing solutions in the literature while promoting the two objectives sought by cloud providers which are ensuring more economies of scale and avoiding tenants hesitation.

The most advantageous solution so far is the Mixed-Tenancy [7]. Our contribution builds on Mixed-Tenancy results and proposes an improvement of multi-tenant SaaS applications while automating instances assignment procedures. Thus, we propose a new artifact called RVC that allows to customize services according to customer requirements. Our contribution concern both services functional customization and automation of the optimal distribution of instances.

This paper treats the architectural part of our RV-Cloud approach and present the different elements of our architectural model. The remainder of this paper is structured as follows. Section 2 identifies the treated problem consisting in instances optimization. Section 3 provides definitions of main notions used in this work as well as introductions to the RV-Cloud approach. Section 4 presents the main contribution of this paper consisting in a rich-variant architecture. Finally, Section 5 is a conclusion of the paper.

## 2. Problem of instances optimization

The emergence of cloud computing has required more and more variability in term of types of services, types of deployment, and cloud participants different roles. Thus, variability modeling is needed to manage the inherent complexity of cloud systems.

SaaS is a delivery model whose basic idea is to provide on-demand client applications on the Internet. SaaS applications are consumed by many customers who have different requirements. Thus, customers who consume the same application generally have different requirements. This type of requirement usually requires alternative software architectures. In other words, when the requirements of the applications are changed, the software architectures of these applications must be adapted to meet them. As a result, the requirements and architectures have intrinsic variability characteristics.

In addition, other problems are raised by Multi-Tenancy which is favored by SaaS to exploit economies of scale. This means that a single instance of an application serves multiple clients. Customers or tenants are for example businesses, clubs or private individuals who have adhered to the use of the application. Even if several clients use the same instance, each one of them feels that the instance is only designated for them. This is archived by isolating tenant data from each other. Unlike single tenancy, Multi-tenancy hosts a plurality of tenants on the same instance.

However, one of the main disadvantages of multi-tenancy applications is the need to ensure the accuracy of all possible configurations of the application in addition to the hesitation of customers to share the infrastructure, the code of the application or data with other tenants. This is because customers are afraid that other tenants may access their data due to a system error, malfunction, or destructive action.

On the other hand, in multi-tenant SaaS applications consumer does not have to worry about doing updates and upgrades, adding security and system patches and ensuring the availability and performance of the service. In addition to this, fast elasticity and pooling of resources are key features of the Cloud [8], which promote variability in the Cloud Computing environment and in particular for multi-tenant contexts.

Operational cost of the application must decrease by sharing computing resources among the plurality of tenants. It is sought to realize a multi-rental application optimized for the operator and the tenant at the same time. For the operator, the cost and effort must be reduced, especially with respect to the use of the IT resource infrastructure. And for the tenant, the data security needs to be improved at the same time. Once a deployment configuration has been created, there will be one or more instances of each deployment level. The deployment configuration is optimal if it generates a minimal cost, using only a minimum number of units of application component instances and the underlying infrastructure layers.

Cloud operators need less infrastructure to offer an application in the MT model than the Single-Tenancy model. But the resources required are not the only way to save costs, the operator can also minimize the effort required to maintain a high number of instances [9].

## 3. User-Aware multi-tenant SaaS approach based on Rich-variant Components

In order to provide a more flexible, more dynamic and more reusable environment for SaaS application providers, our approach proposes a user-aware tenancy based on the use of RVC.

### 3.1 Stakeholders

As a first step, we start by defining the different stakeholders involved in our problem. We distinguish three different stakeholders: the SaaS Provider, the Customer / Tenant, and the End User. Their definitions are as follows:

**SaaS Provider:** A SaaS provider is a company that develops an application and provides it to the market. A SaaS operator deploys, runs, and maintains applications on a rented or owned hardware infrastructure. In our work, we consider the SaaS provider and the SaaS operator as the same entity.

**Customer / Tenant:** A Customer or a Tenant, or even a Tenant Customer, is a company that pays to use an application provided by the SaaS provider. The term Customer is used in a commercial point of view. The technical term is Tenant. For the rest of our work, both terms will be used interchangeably as they refer to the same entity.

**End User:** An End User is a person or employee who has the access to an application and, therefore, interacts with it. Each end user belongs to the staff of exactly one customer / tenant or is employed by exactly one customer / tenant.

When designing an application, the application provider predefines end-user profiles categorizing the business needs of different end-users according to their missions.

### 3.2 Rich-Variant Component

In a second step, we define the concept of RVC component on which our approach is based. The definition of an RVC depend on the definitions of a software component existing in the literature. One of the first definitions of the component concept was proposed by Booch [8] which defines a reusable software component by " *a logically cohesive, loosely coupled module that denotes a single abstraction.*" Besides, one of the most quoted and globally accepted definitions is given by Szyperski [9] who defines a software component as " *a unit of composition with contractually specified interfaces and explicit context dependencies only. A software component can be deployed independently and is subject to composition by third parties.*"

Indeed, depending on theese definitions of a software component, the following definition is a definition of what is called an RVC component in our present work.

**Rich-Variant Component:** An RVC is defined as an application building block that encapsulates an atomic functionality. All functionalities and properties that the RVC provides to and requires from other RVCs must be captured by a described interface, through which all interactions flow. In addition, an RVC has several deployment variants, it can be used in its different ways and therefore changes behavior dynamically depending on the functionality and the end user. Moreover, it is very important for this work, that the RVCs can be deployed independently of each other.

In fact, the focus on the possibility of independent deployment is of particular importance to our work. This is because one of the main challenges is that RVCs are deployed multiple times, to be used by different tenants. This is only possible if they can be separated from each other.

In our approach, SaaS applications are built from a number of basic RVCs, each RVC provides an atomic functionality and dynamically changes behavior depending on the available end user profile. Our SaaS applications built from RVCs then behave differently depending on the end user profile available.

### 3.3 Introduction to the RV-Cloud approach

Through our work, we seek to exploit economies of scale while avoiding the problem of customer hesitation to share with others as well as allowing better communication between customer communities.

Our approach proposes a provider platform from which information is exchanged between the provider and his customers. The provider presents his offers and the customers express their needs and requirements.

In addition to collecting customers functional requirements, the main idea of our work is to collect even the deployment sharing requirements. This allows to consider deployment requirements of all tenants when calculating an optimal distribution of application instances over customers renting this application. The following are the definitions of what is called functional requirement, deployment requirement, and optimal distribution in our present work.

**Functional Requirement:** A functional requirement consist in a selection of application functionalities based on variation points proposed by the application provider.

**Deployment Requirement:** A Deployment requirement is a description of a customer's desire or unwillingness to share a part of the application. It is necessary for a tenant to provide a number of deployment requirements for the deployment of an entire application.

**Optimal Distribution:** It's about a distribution of application instances on its tenant customers. A distribution must necessarily meet the functional requirements and deployment requirements defined by all tenants. This distribution is optimal if it results a minimal cost using an optimal number of RVC instances.

## 4. Our Rich-Variant Architecture

The overall vision of the architecture of our approach is presented in Figure 1. The main elements of our architecture are the configurable applications, the Variability Engine, the Execution Framework, and the Optimal Distribution. In the following subsections, we will explain and detail each element of our Rich-Variant architecture.

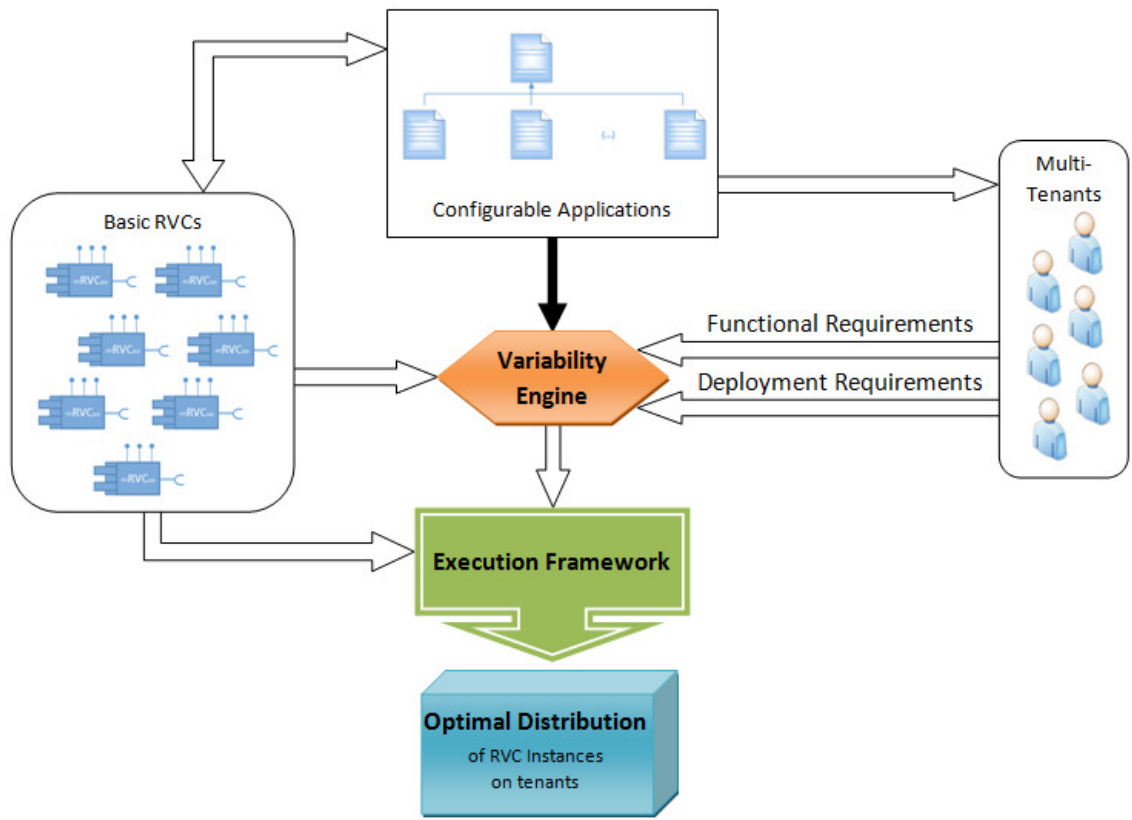

Fig. 1 Our Rich-Variant Architecture.

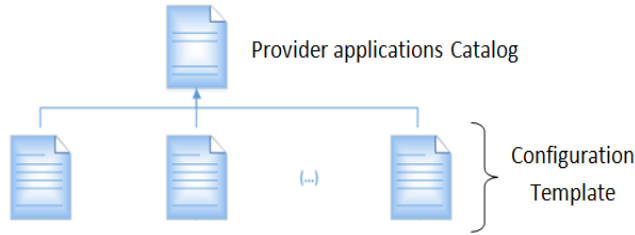

Fig. 2  Organization of applications.

### 4.1 Configurable Applications

The first element of our architecture is the applications offered by the provider. These are applications built based on RVCs. The applications offered by the provider are organized as shown in Figure 2.

In a first level, the highest level of abstraction, we have the provider Catalog which is a formal description of all applications available and offered by this provider. The Catalog presents applications functional variability by describing the different functionalities of each application in addition to the specification of the points of variability, thus showing to the customers how each application can be customized.

Considered as an instantiation of the Catalog related to an application, the Configuration Template comes in a second level of abstraction describing the basic RVCs that must be linked to create and build a given application. It describes the RVCs and their variants needed to realize the different functionalities of the given application.

The last abstraction level is represented by the Rich-Variant Configuration generated by the Variability Engine that is the subject of the next section.

### 4.2 Variability Engine

All tenants use the same Variability Engine that captures the functional requirements and deployment requirements of each tenant. The Variability Engine generates Rich-Variant Configurations which are each specific to a tenant.

Generated from a given Configuration Template, a Rich-Variant Configuration describes a specific application tailored to the needs of a specific tenant with behavior that dynamically changes when executed according to the end-user's point of view available. At this level, the values of the parameters or points of variability of each RVC are defined, it is the functional description of the concrete application that will be provided to the tenant. A Rich-Variant Configuration is derived based on the functional requirements of a specific tenant.

As we have already mentioned, our SaaS applications are built of RVCs. Each RVC has a number of variants. And every application functionality is achieved through the use of a number of variants of the RVCs building the application.

From our platform, tenants view the provider Catalog, choose the functionalities they want to have in an application, and specify their deployment requirements for each functionality in the application.

An example of a deployment requirement is "I do not want to share functionality F with any other tenant", or "I want to share functionality F with tenant X" ...

To facilitate the collect of deployment requirements, we formalized their expressions by defining four possible cases. Tenants can express their deployment requirements concerning each application functionality using the following expressions:

- *SWAny: Share with anyone (default value)*
- *SWJ(X): Share with just X ;*
- *DSW(X): Don't share with X ;*
- *DSWAny: Don't share with anyone.*

Where X can take the values: "P" (as Partners), "Cp" (as Competitors), "Ti" (for a specific Tenant), or a list of the previous values.

Requirements are ordered in a table where are stored requirements of each tenant for each application functionality. We have a such table for each application. When a tenant does not specify deployment requirement for a functionality, it means that the tenant has no problem sharing this functionality. In this case, we take the default value which means "Share with any other tenant".

On the side of customers or tenants, we talk about sharing functionalities, whereas on the side of providers, we talk about sharing variants of RVCs. As a result, the final step of the Variability Engine is to translate customer requirements concerning functionalities into requirements concerning variants of RVCs. Two tenants can not share a functionality means that they can not share variants of RVCs that participate in the realization of this functionality. Then we get one table by RVC containing each tenant requirements for each RVC variant. However, there may be several expressions in one table cell, to settle this problem we apply the transition rules presented in Table 1, where Z can take one of four possible expressions (ie, whatever Z).

Table 1: Transition Rules

| 1st expression | 2nd expression | Combination result |
|---|---|---|
| SWA | Z | Z |
| DSWA | Z | DSWA |
| DSW(X) | DSW(Y) | DSW(X,Y) |
| SWJ(X) | SWJ(Y) | DSWA |
| DSW(X) | SWJ(Y) | SWJ(Y) |
| DSW(X) | SWJ(X) | DSWA |
| SWJ(0) |  | DSWA |
| DSW(0) |  | SWA |

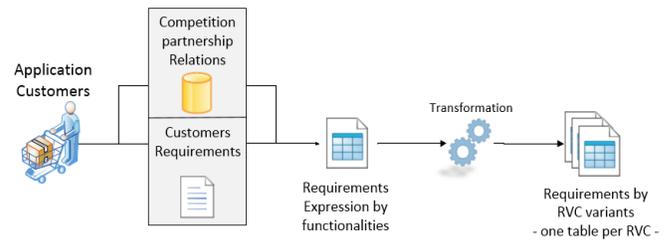

Fig. 3  Variability Engine functioning.

The Variability Engine captures tenants' functional requirements as well as tenants' deployment requirements. It handles the data to give both tenant-specific Rich-Variant Configuration and tables of requirements concerning variants of RVCs, one table for each RVC. Each RVC variant-ordered table is the input of our Execution Framework. Figure 3 schematizes the Variability Engine treatment.

### 4.3 Execution Framework

Our Execution Framework takes as input the ordered requirements of application tenants provided by the Variability Engine, and it gives as output the Optimal Distribution of application instances on tenants of the application. The work of the Execution Framework with the progress of its various steps is shown in Figure 4. The Execution Framework reproduces the treatment for each RVC.

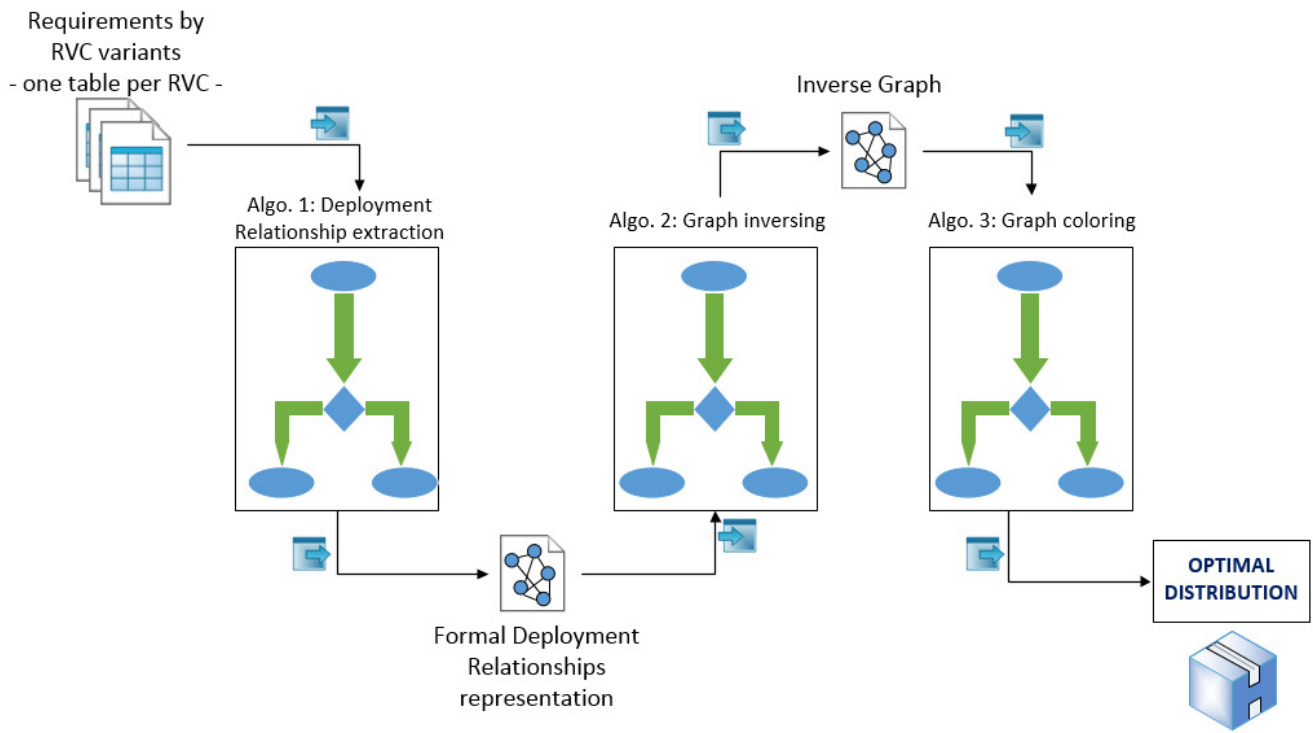

Fig. 4  Description of our Execution Framework.

The first steps of the Execution Framework treatment consist in the execution of a first algorithm that aims to extract the deployment relationships. Indeed, based on the RVCs variants-ordered requirements of tenants, the algorithm retrieves the deployment relationship between tenants concerning one RVC as a formal representation based on graphs.

**Deployment relationships:** Deployment relationships concerning an RCV are created based on all deployment constraints and requirements identified by all customers. Deployment relationships describe which tenants can share which variants of the RVC. In our work, we formally represent deployment relationships with graphs, one graph by an RVC.

For the formal representation of deployment relationships, we work with Undirected Edge Labeled Graphs. Indeed, while vertices represent tenants, edges represent if two tenants can share variants or not. Besides, labels on edges indicate the variants involved in sharing relationship represented by the edge. When an edge has no label, that means that sharing relationship concerns the RVC with all its variants.

An example of deployment relationships representation based on an Undirected Edge Labeled Graph is presented in Figure 5. It's about deployment relationships of six tenants T1 to T6 concerning an RVC having four variants A, B, C, and D.

The second Execution Framework treatment step consists in executing a second algorithm that aims to inverse the graph of deployment relationships provided as an output by the first step to have the inverse graph of deployment relationships.

The third Execution Framework treatment step consists in the execution of a third algorithm -Algorithm A from Figure 6- that colors the input graph. This algorithm takes as input the inverse graph of the deployment relationships. The coloring of the inverse graph according to our coloring algorithm makes it possible to deduce the optimal distribution of the instances of the RVC variants on tenants of the application.

```
Algorithm A: The Coloring Algorithm
----------------------------------------------------------------
Input : m number of tenants T1, ...,Tm,
        and n number of RVC variants V1, ..., Vn
Output : C ={C1, ..., Cd}, with d number of colors used
----------------------------------------------------------------
1:  Give a color to T1 (for all variants) and  put d=1
2:  For i from i=2 to i=m do
3:      For j from j=1 to j=n do
4:          For k from k=1 to k=d do
5:              if Ti is not adjacent to any T from Ck according to Vj
6:              then give the color Ck to Ti.Vj
7:              and put j=j+1
8:              else   if  k=d
9:                     then put d=d+1
10:                    and give the new color Cd to Ti.Vj
11:                    and put j=j+1
12:                    else put k= k+1
13:                    end if
14:         end if
15:     end For
16:   end For
17: end For
18: return C ={C1, ..., Cd}
```

Fig. 6  Our coloring function's algorithm

While completing these steps, our Execution Framework achieves its goal by providing application tenants with application instances deduced from optimal distributions given as output of the algorithm for each RVC building the application.

The different algorithms used in the three steps of the treatment of our Execution Framework as well as the types of graphs used was detailed in previous work [10] while mentioning their origins and the main idea of their use.

### 4.4 Optimal Distribution

Let's first recall the definition of an Optimal Distribution cited in the previous sections:

**Optimal Distribution:** It's about a distribution of application instances on its tenant customers. A distribution must necessarily meet the functional requirements and deployment requirements defined by all tenants. This distribution is optimal if it results a minimal cost using an optimal number of RVC instances.

Since our SaaS applications are built from a number of RVCs, calculating the Optimal Distribution of instances of an application will then entails calculating the optimal distribution of instances of RVCs building the application. A big part of our contribution is a treatment that recurs on every RVC building the application. What we need for this RVC treatment is the deployment relationships concerning each RVC resulting from the translation of tenants requirements for functionalities and which indicates for each two tenants whether or not they can share a specific variant of an RVC.

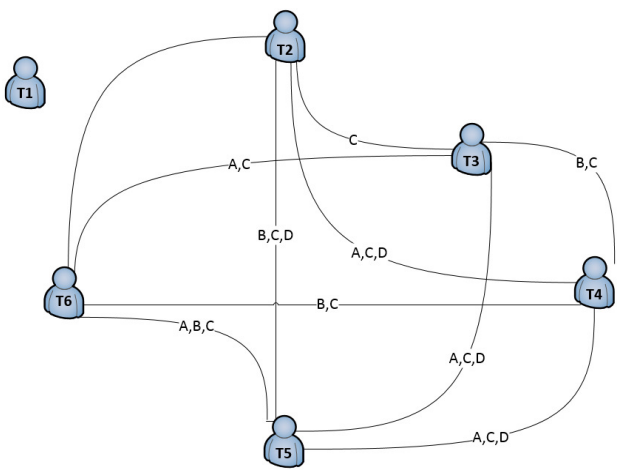

Fig. 5  Example of deployment relationships graph-based representation

The Optimal Distribution of RVC-based application instances on tenants is derived from the optimal distributions of variants instances of each RVC on tenants while meeting their requirements.

## 5. Conclusions

Flexibility, dynamicity, and reusability are challenging issues for cloud environments and particularly for SaaS application providers. Therein, our user-aware multi-tenant SaaS approach called RV-Cloud approach comes to create a more flexible, more dynamic, and more reusable SaaS environment while using RVCs. In this context, this paper treats the conceptual part of our RV-Cloud approach and present the different elements of our conceptual model. After identifying the treated problem in our work consisting in instances optimization in cloud computing environments, the paper provided definitions of main notions used as well as introductions to our RV-Cloud approach. Later, we get to present the main contribution of this paper consisting in our rich-variant architecture. As future work, we think about projecting our approach in the domain of Model-driven engineering for a more modern and more general vision.

**Houda Kriouile** was born at Nancy in France, in 1990. She is a Ph.D. candidate, and a member of the IMS team from the SIME laboratory, National Higher School for Computer Science and Systems Analysis (ENSIAS), Mohammed V University of Rabat, Morocco. She received her engineer degree in computer science and software engineering in 2012 from the ENSIAS School at Mohammed V University of Rabat, Morocco.

**Bouchra El Asri** is a professor in the Software Engineering Department, National Higher School for Computer Science and Systems Analysis (ENSIAS), Mohammed V University of Rabat, Morocco. She received her Ph.D. degree in computer Science from National Higher School for Computer Science and Systems Analysis (ENSIAS). Her research activities focus on Cloud Computing, Engineering of complex systems based on multi-dimensional components, Generation of safe components, Development of dynamic systems based on contextual services, Model Driven Engineering, Transactional services, Specific domain component engineering.